\documentclass[11pt]{article}
\pdfoutput=1
\begin{document}
\title{Donut-shaped Bubbles Formed by Raindrops}
\author{Marc Buckley, Florian Bernard, Fabrice Veron \\
\\\vspace{6pt} College of Earth, Ocean, and Environment, \\ University of Delaware, Lewes, DE 19958, USA}
\maketitle
\section{Video description}
Single free-falling freshwater drops were generated with no initial velocity by hypodermic needles, at an altitude of 3.61 m above a still freshwater surface. High resolution high speed videos (0.13 mm/pixel, 500 frames/second) of the dynamics of the impact were acquired. A few milliseconds after forming the usually observed cavity, canopy and coronet (Prosperetti et al. 1993\footnote[1]{A. Prosperetti and H. N O\~{g}uz, "The impact of drops on liquid surfaces and the underwater noise of rain", Annu. Rev. Fluid Mech. 25:577-602 (1993)}), drops of diameters typically greater than 3.8 mm consistently generated toroidal (donut-shaped) air bubbles upon impact at the water surface. Videos of the dynamics of the impingement were successively taken from different angles, and with a 105 mm lens focused on different regions of the event. These allowed for a qualitative description and hypothetical explanation of the observed phenomena, which are presented alongside the actual video footage, in the hereby displayed fluid dynamics video.\\
The video displays a cartoon of the experimental setup, followed by two simultaneously running videoclips of the drop impact upon the water surface, from 2 different viewpoints (front and top). As the event occurs over less than 1 second, the videos were slowed down to a frame rate of 12 frames/sec. The footage suggests that the air is trapped by the converging rim (or coronet) of the closing canopy. The very bottom of the cavity then rises rapidly, and collides with the falling water mass provided by the fully converged canopy rim. This incidentally prevents the formation of a secondary jet. The center (or donut hole) of the unstable toroidal air bubble eventually moves outward, thus yielding a hemispherical air bubble.

\end{document}